\documentclass[aps,prl,twocolumn,superscriptaddress,showpacs]{revtex4}

\usepackage{graphicx}
\usepackage{color}

\begin{document}

\title{Logarithmic growth law in the two-dimensional Ising spin glass state resulting from the electron doping in single-layered manganites}

\author{R. Mathieu\cite{newaddress}}
\affiliation{Spin Superstructure Project (ERATO-SSS), JST, AIST Central 4, Tsukuba 305-8562, Japan}

\author{J. P. He}
\affiliation{Spin Superstructure Project (ERATO-SSS), JST, AIST Central 4, Tsukuba 305-8562, Japan}

\author{Y. Kaneko}
\affiliation{Spin Superstructure Project (ERATO-SSS), JST, AIST Central 4, Tsukuba 305-8562, Japan}
\affiliation{ERATO multiferroics project, JST, c/o Department of Applied Physics, University of Tokyo, Tokyo 113-8656, Japan}

\author{H. Yoshino}
\affiliation{Department of Earth and Space Science, Faculty of Science, Osaka University, Toyonaka, 560-0043 Osaka, Japan}

\author{A. Asamitsu}
\affiliation{Spin Superstructure Project (ERATO-SSS), JST, AIST Central 4, Tsukuba 305-8562, Japan}
\affiliation{Cryogenic Research Center (CRC), University of Tokyo, Bunkyo-ku, Tokyo 113-0032, Japan}

\author{Y. Tokura}
\affiliation{Spin Superstructure Project (ERATO-SSS), JST, AIST  Central 4, Tsukuba 305-8562, Japan}
\affiliation{ERATO multiferroics project, JST, c/o Department of Applied Physics, University of Tokyo, Tokyo 113-8656, Japan}
\affiliation{Correlated Electron Research Center (CERC), AIST Central 4, Tsukuba 305-8562, Japan}
\affiliation{Department of Applied Physics, University of Tokyo, Tokyo 113-8656, Japan}

\date{\today}

\pacs{75.50.Lk, 75.47.Lx, 75.40.Gb}

\begin{abstract}

The ac-susceptibility of the electron doped single-layered manganite La$_{1.1}$Sr$_{0.9}$MnO$_4$ is analyzed in detail. A quasi  two-dimensional (2$D$) antiferromagnetic (AFM) order with Ising anisotropy is stabilized below $T_N$ $\sim$ 80K. We show that below $T_N$, a rare 2$D$ spin-glass (SG) correlation develops with the same Ising anisotropy as the AFM state. Using simple scaling arguments of the droplet model, we derive a scaling form for the ac-susceptibility data of a 2$D$ SG, which our experimental data follows fairly well. Due to simplifications in this 2$D$ case, the proposed scaling form only contains two unknown variables $\psi\nu$ and $\tau_0$. Hence, the logarithmic growth law of the SG correlation predicted by the droplet model is convincingly evidenced by the scaling of our experimental data. The origin and nature of this 2$D$ SG state is also discussed.

\end{abstract}

\maketitle

Depending on the size of the ionic radii of the $R^{3+}$ and $A^{2+}$ ions in the $(R_{1-x}A_x)_{n+1}$Mn$_n$O$_{3n+1}$ manganites ($R$ is a rare earth and $A$ is an alkaline earth element), the ferromagnetic metallic phase, or the charge- and orbital-ordered (CO-OO) can be stabilized\cite{orbital}. It has been shown in the three-dimensional perovskite case ($n$=$\infty$, $R_{1-x}A_x$MnO$_3$) that in the presence of the quenched disorder introduced by the solid solution of the $R^{3+}$/$A^{2+}$ ions, the long-range CO-OO phase collapses in a first-order manner\cite{tomioka-diag}, and only a nanometer-sized CO-OO correlation, and associated spin glass (SG) state remain\cite{roland-ebmo}. In the two-dimensional case ($n$=1, $R_{1-x}A_{1+x}$MnO$_4$), the short-ranged CO-OO state, or CE-glass\cite{dagotto}, is also observed, albeit the first-order like collapse of the CO-OO does not occur\cite{roland-diag}. The associated SG state is atomic-like in both cases\cite{roland-ebmo,roland-esmo}, as it results of the uniform fragmentation of the ferromagnetic zig-zag chains constituting the CO-OO structure on the nanometer scale\cite{roland-diag}. This ``orbital-master/spin-slave'' relationship is valid for hole-doped manganites with half-doping or larger (0.5 $\leq$ $x$ $<$ 1) \cite{aka-new,yu}. However in hole-underdoped manganites, a low-temperature SG state may appear even if the CO-OO is long ranged\cite{yu,roland-pcmo}, as magnetic frustration is introduced with the ``excess electrons'' in the structure.

We here investigate the spin-glass state of electron-doped single-layered manganites (i.e. with $x<0$). While the undoped LaSrMnO$_4$ ($x$=0) is antiferromagnetic, an uncommon two-dimensional (2$D$) SG state with Ising anisotropy is observed in the electron doped La$_{1.1}$Sr$_{0.9}$MnO$_4$. Using simple scaling arguments from the so-called droplet model, we derive a simple form of the full dynamical scaling of the temperature- and frequency-dependent ac-susceptibility of such a 2$D$ spin-glass, which is found to describe our experimental data. Due to the great simplification of the scaling form in this 2$D$ case, our scaling analysis convincingly confirms the logarithmic growth law of the spin-glass correlation predicted by the droplet model. The origin of this 2$D$ Ising SG, as well as the relationship between orbital and spin degree of freedom are also discussed.

High quality single crystals of the La$_{1.1}$Sr$_{0.9}$MnO$_4$ were grown by the floating zone method. The phase-purity of the crystals was checked by x-ray diffraction. The ac-susceptibility $\chi$($\omega=2\pi f$) data was recorded as a function of the temperature $T$ and frequency $f$ on a MPMSXL SQUID magnetometer equipped with the ultra low-field option (low frequencies) and a PPMS6000 (higher frequencies), after carefully zeroing or compensating the background magnetic fields of the systems. Additional phase corrections were performed for some frequencies. The heat capacity $C$ was recorded using a relaxation method on the same measurement system.


LaSrMnO$_4$ ($x$=0) is an antiferromagnet (AFM) below $T_N$ $\sim$ 120K, with the K$_2$NiF$_4$ structure\cite{larochelle, asam}. In these single-layered crystals with the typical lattice-parameter ratio $c/a$ $>$ 3, the MnO$_2$ planes ($ab$-planes) are isolated by blocking (La/Sr)$_2$O$_2$ layers, yielding a 2$D$ Mn-O network which limits the spatial extent of the magnetic and charge- and orbital correlation along the $c$-axis of the structure. All Mn ions are Jahn-Teller Mn$^{3+}$ ions with three $t_{2g}$ electrons ($d_{xy}$, $d_{yz}$, and $d_{zx}$ orbitals), and an $e_g$ electron occupying the $c$-oriented $d_{3z^2-r^2}$ orbital\cite{z2}; the last 3$d$ orbital, the $ab$-oriented $d_{x^2-y^2}$, is unoccupied. As a result, the AFM state shows an Ising anisotropy, with spins parallel to the $c$-axis of the structure\cite{larochelle,asam}. When electrons are added to LaSrMnO$_4$ by increasing the amount of La$^{3+}$ with respect to Sr$^{2+}$, as in La$_{1.1}$Sr$_{0.9}$MnO$_4$, Mn$^{2+}$ ions are created, with both $d_{3z^2-r^2}$ and $d_{x^2-y^2}$ orbitals occupied. 
\begin{figure}
\includegraphics[width=0.46\textwidth]{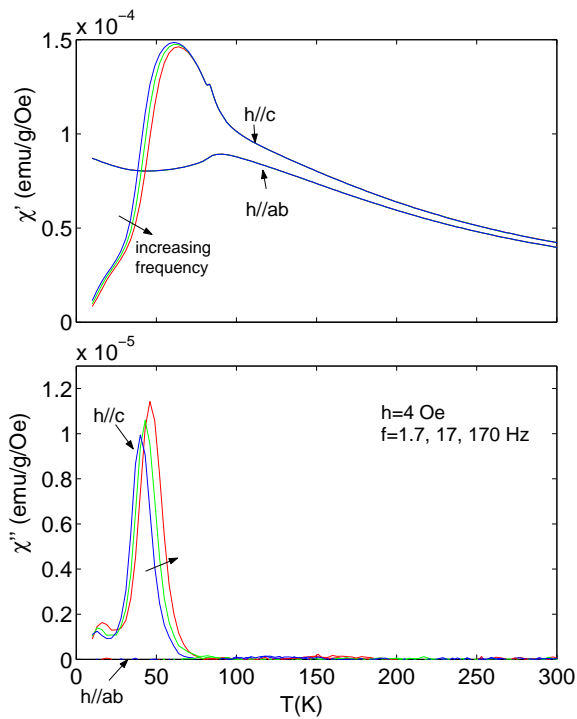}
\caption{(Color online) Temperature $T$ dependence of the in-phase (upper panel) and out-of-phase (lower panel) components of the ac-susceptibility $\chi'(T,\omega = 2\pi f)$ and $\chi''(T,\omega)$ for different orientations of the probing field; $h$ (field amplitude) = 4 Oe and $f$ (frequency) = 170, 17, and 1.7 Hz.}
\label{fig-XT}
\end{figure}

\begin{figure}
\includegraphics[width=0.46\textwidth]{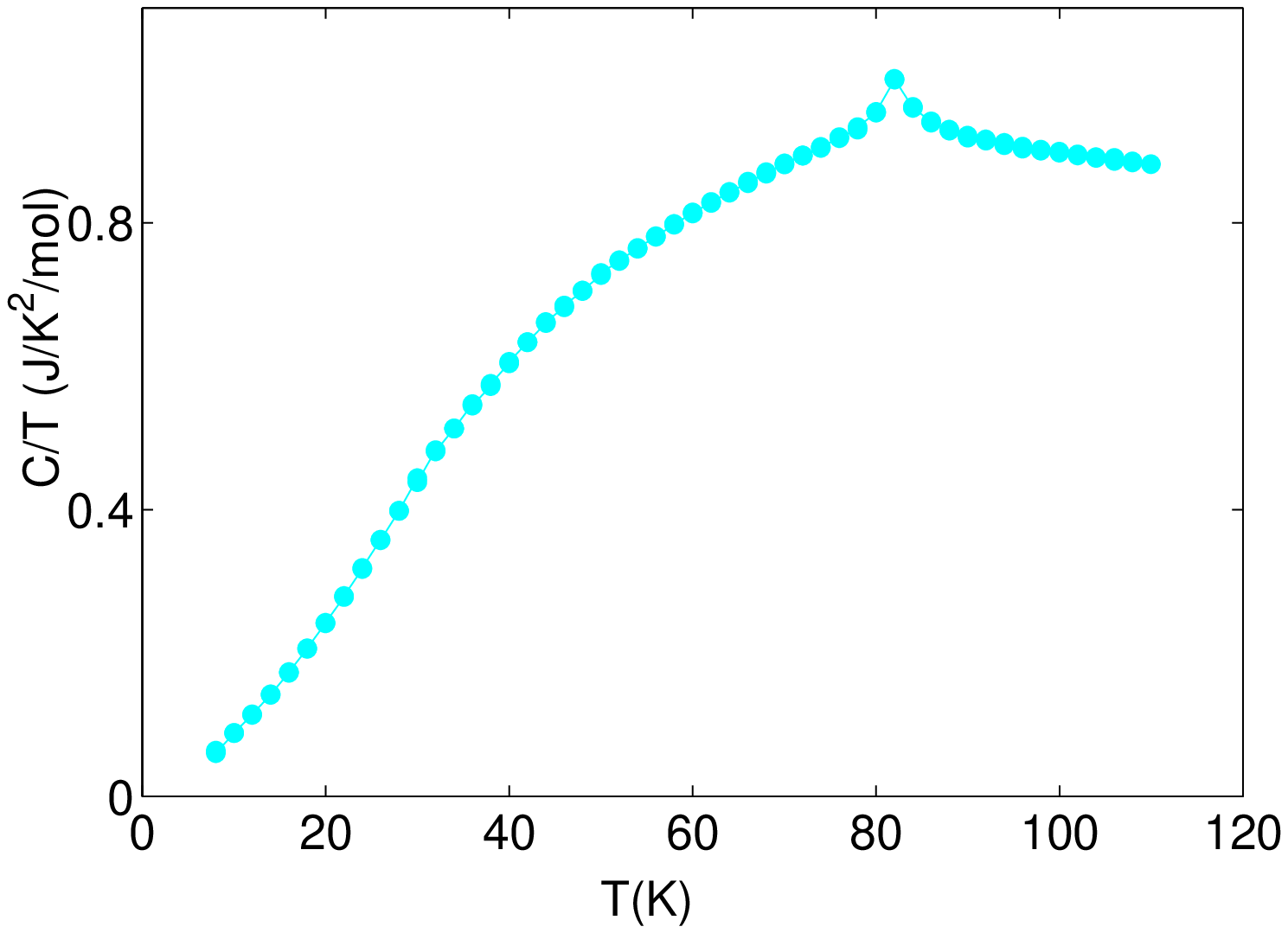}
\caption{(Color online) Temperature $T$-dependence of the heat capacity $C$, plotted as $C/T$.}
\label{fig-HC}
\end{figure}

As seen in the upper panel of Fig.~\ref{fig-XT}, the in-phase component of ac-susceptibility $\chi'$($\omega$,$T$) recorded with a probing magnetic field $h$ applied along the $c$-axis of the structure, exhibits two overlapping peaks at low temperatures. A sharp peak near 80 K, and a broad peak above 50 K with large frequency dependence, nearly masking the former. Instead, $\chi'$($\omega$,$T$) recorded with $h$ within the $ab$-plane is rather flat below 80K, and frequency independent. Accordingly, the corresponding out-of-phase component, $\chi''$($\omega$,$T$) is negligible at all temperatures (see lower panel of Fig.~\ref{fig-XT}). The temperature dependence and anisotropy of $\chi'$($\omega$,$T$) are reminiscent of those of an AFM with spins parallel to the $c$-axis, albeit with a broad frequency dependent peak superposed in the $c$-direction. The AFM transition is confirmed by heat capacity measurements, as seen in Fig.~\ref{fig-HC}. Below $T_N$, the heat capacity only exhibit a broad feature, which may reflect magnetic disorder or glassiness at low temperatures\cite{HC}. The low-temperature broad peak observed in the $\chi'$($\omega$,$T$) displays the same Ising anisotropy as the AFM state, and $\chi'$($\omega$,$T$) recorded with $h // c$ resembles that of SG or superparamagnets\cite{roland-ebmo,fullscal3d}. The additional $e_g$ electrons in Mn$^{2+}$ are likely to hop onto neighboring empty  $e_g$ states and host the ferromagnetic (FM) interaction by double-exchange mechanism\cite{orbital}, yielding the appearance of FM correlation in the original AFM structure of LaSrMnO$_4$. Hence a reentrant SG state, or a superparamagnetic state might appear in La$_{1.1}$Sr$_{0.9}$MnO$_4$. The Hund coupling on Mn$^{2+}$ sites align the spins of the $e_g$ and $t_{2g}$ orbitals along the $c$-axis, yielding the observed Ising anisotropy of the low-temperature disordered state. 

\begin{figure}
\includegraphics[width=0.5\textwidth]{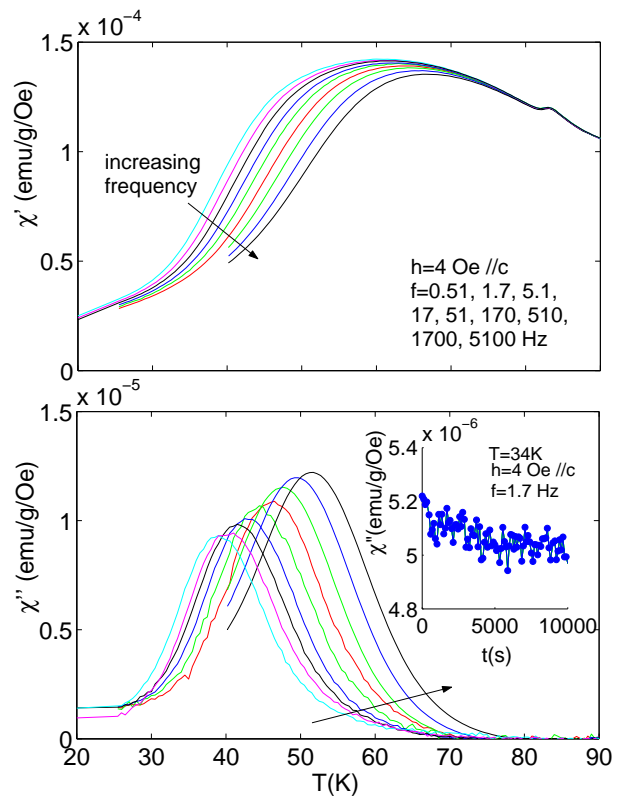}
\caption{(Color online) Temperature dependence of the in-phase (upper panel) and out-of-phase (lower panel) components of the ac-susceptibility $\chi'(T,\omega)$ and  $\chi''(T,\omega)$ for different frequencies. The probing field is applied along the $c$-direction. The inset shows the relaxation of $\chi''(T,\omega)$ as a function of time $t$ after a quench from high temperatures.}
\label{fig-XT2}
\end{figure}

We now study the time and frequency dependence of the ac-susceptibility in more detail, in order to determine the nature of the low-temperature state. Because of the Ising anisotropy, the probing ac magnetic field is applied along the $c$-axis. Three-dimensional (3$D$) SG states have frequently been observed in similar single-layered\cite{roland-esmo,roland-pcmo,roland-diag}, as well as in pseudo-cubic perovskite\cite{roland-ebmo} manganites doped with holes. It could thus be expected that a similar SG state appears in the present electron-doped case. As seen in the lower inset of Fig.~\ref{fig-XT2}, La$_{1.1}$Sr$_{0.9}$MnO$_4$ displays aging features characteristic of spin glasses\cite{ghost,roland-ebmo,roland-pcmo}, reflecting the slow rearrangement of the spin configuration toward its equilibrium state at a given temperature within the glassy phase after a quench to that temperature\cite{ghost}.

In each $\chi''$($\omega$,$T$) curve in the lower panel of Fig.~\ref{fig-XT2}, we can define a frequency dependent freezing temperature $T_f$($f$), below which the system is out-of-equilibrium (typically the temperature onset of $\chi''$($\omega$,$T$), see below). In each measurement, the system is probed on a typical time scale $t_{obs}$ $\sim$ 1/$\omega$ (= 1/2$\pi f$) or a length scale $L$(1/$\omega$). One can check whether the dynamical slowing down toward the SG phase transition occurs by scaling $t_{obs}$ = $\tau$($T_f$) with the reduced temperature $\epsilon=(T_f(f)-T_g)/T_g$ ($T_g$ is the spin glass phase transition temperature) using the power law form\cite{ghost,roland-ebmo}: $$\frac{\tau}{\tau_0} = \epsilon^{-z\nu}$$ where $z$ and $\nu$ are critical exponents, and $\tau_0$ the flipping time of the fluctuating entities (typically $\tau_0$ $\sim$ 10$^{-13}$ s for atomic spins). No perfect scalings of the $T_f$($f$) data can be obtained, and the best scaling implies unphysical $z\nu$ values exceeding 30, largely dependent on the choice of $T_f$. Hence the system does not seem to undergo a three-dimensional (3$D$) SG phase transition. The frequency dependence of the $\chi'$($\omega$,$T$) and $\chi''$($\omega$,$T$) curves shown in Fig.~\ref{fig-XT2} are rather broad, with the onset of $\chi''$($\omega$,$T$) much less sharper than in the case of archetypal SG\cite{ghost,roland-ebmo,roland-esmo}. Thus, in order to remove uncertainties related to the determination of $T_f(f)$, one can perform a so-called full scaling of the susceptibility data. In the case of 3$D$ SG, the full dynamical scaling can be written as: $\chi''T\epsilon^{-\beta}=F(2\pi f \tau_0\epsilon^{-z\nu})$, where $\beta$ is another critical exponent and $F$ a numerical function\cite{fullscal3d}. However, no scaling characteristics of the $\chi''$($\omega$,$T$) curves (with $T$ $>$ $T_g$) can be obtained for any sets of $\beta$, $z\nu$, $T_g$, and $\tau_0$ parameters.

The frequency dependence of $\chi''$($\omega$,$T$) is similar in its broadness to that of superparamagnets\cite{fullscal3d}; however superparamagnets do not exhibit aging or other glassy features\cite{fullscal3d,ghost}. The scaling of the $T_f$($f$) data to $$log(\frac{\tau}{\tau_0})\propto \frac{1}{T_f(f)}$$ fails equally (and $\tau_0$ $>$ 10$^{-19}$ s), implying that our data does not reflect the superparamagnetic relaxation\cite{superparam}. Due to the simple Arrhenius law relating $T_f$ and $t_{obs}$, the superparamagnetic relaxation implies for $\chi''$($\omega$,$T$): $\chi''=F[-Tlog(2\pi f\tau_0)]$ where $F$ is a numerical function\cite{superparam}. The scaling of our experimental data using this scaling form fails (which can be expected as the magnitude of the maximum of $\chi''$($\omega$,$T$) depends on the frequency), and yields unphysical $\tau_0$ values, as in the case of the simple $T_f$($f$) scalings.

\begin{figure}
\includegraphics[width=0.46\textwidth]{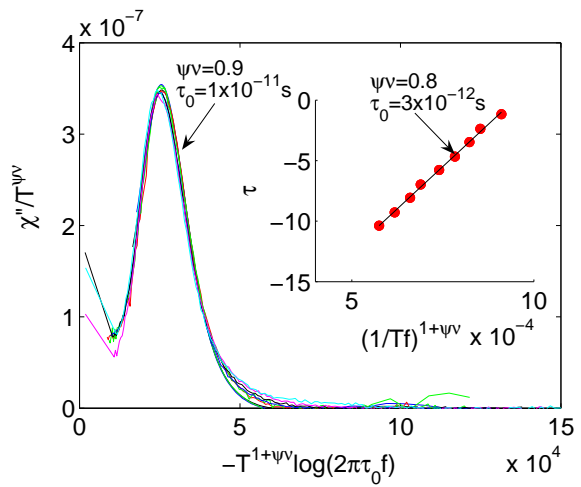}
\caption{(Color online) Full scaling of the $\chi''(T,\omega)$ curves using the scaling law derived in the main text. $\psi$ and $\nu$ are critical exponents while $\tau_0$ reflects the flipping time of the fluctuating entities. The inset shows the scaling of the observation time $\tau=1/\omega$ (as $log(\tau)$) with $1/T_f^{1+\psi\nu}$; $T_f$($\omega$) is the frequency dependent freezing temperature.}
\label{fig-scal}
\end{figure}

We now check whether the disordered state observed at low temperatures in La$_{1.1}$Sr$_{0.9}$MnO$_4$ behaves like a 2$D$ SG, i.e. with 2$D$ SG correlation developing below $T_N$. In that case, the dynamical slowing down is expressed with the generalized Arrhenius law ($T_g$=0): $$log(\frac{\tau}{\tau_0})\propto \frac{1}{T_f(f)^{1+\psi\nu}}$$ where $\psi$ is the energy barrier exponent\cite{per,droplet}. As shown in the inset of Fig.~\ref{fig-scal}, the scaling of the $T_f$($f$) data works fairly well, with meaningful values for $\psi\nu$ and $\tau_0$. The obtained value of $\tau_0$ $\sim$ 10$^{-12 \pm 1}$ s is close to the microscopic spin flipping time which indicates that the SG state is nearly atomic, with fluctuating entities of the nanometer order. The dynamical exponent $\psi\nu$ value of 0.8  $\pm$ 0.1 is reasonably smaller than 1\cite{per}, and similar to those of other 2$D$ systems. Using a similar analysis, $\psi\nu$ $\sim$ 0.9 was obtained in (Cu,Mn)/Cu multilayers with 2$D$ SG character\cite{per}, while $\psi\nu$ $\sim$ 1.5 was obtained for B$_{12}$ cluster compound HoB$_  {22}$C$_2$N\cite{mamiya}. Hence the low-temperature state La$_{1.1}$Sr$_{0.9}$MnO$_4$ seems to behave like a 2$D$ SG. However, as in the previous 3$D$ and superparamagnetic $T_f$($f$) scalings, $T_f$ may not be accurately defined, and we need to perform a full scaling analysis. For that purpose, we derive a scaling form\cite{note2} for the dynamical slowing down in a 2$D$ SG, based on the droplet picture.

The generalized Arrhenius law mentioned above stems from the droplet model\cite{droplet}, and the fact that thermal activation over energy barriers $L^\psi$ for a droplet excitation of length $L$ implies the logarithmic growth\cite{droplet,per}: 
$$L \propto (Tlog(t))^{1/\psi}$$
This thus implies that $L$ compares to the correlation length $\xi$ as: 
$$\frac{L}{\xi} \propto (\frac{Tlog(t)}{\xi^\psi})^{1/\psi}$$
$\frac{Tlog(t)}{\xi^\psi}$ is thus the natural scaling variable for the spin autocorrelation function $q(t)$\cite{dynscal}, as 
$$q(t)=t^{-x}F(\frac{Tlog(\frac{t}{t_0})}{\xi^\psi})$$
 $F$ is a functional form and $x$ an exponent. In two dimensions, $T_g$=0, $x$=$\frac{1}{2}(d-2+\eta)$=0 and $\xi$ diverges as $T^{-\nu}$, yielding the fairly simple form:
$$q(t)=F(T^{1+\psi\nu}log(\frac{t}{t_0}))$$
 $q(t)$ is related to the time dependent zero-field susceptibility $\chi(t)$ and the experimental susceptibility $\chi$($\omega$,$T$) by the fluctuation-dissipation theorem as: $\chi(t)=\frac{1-q(t)}{T}$ and $\chi''(\omega)=-\frac{\pi}{2}\frac{d\chi(t)}{dlog(t)}$ \cite{lundgren}. Hence combining these expressions, we can write: $\chi''(\omega=2\pi f,T)=-\frac{\pi}{2}T^{\psi\nu}F'(T^{1+\psi\nu}log(\frac{\tau}{\tau_0}))$, which can be rewritten, using $G$=$F'$: 
$$\chi''(2\pi f,T)T^{-\psi\nu}=G(-T^{1+\psi\nu}log(2\pi \tau_0 f))$$
This scaling form is quite simple, with only two parameters $\psi\nu$ and $\tau_0$, due to the above mentioned simplifications in the 2$D$ case. Our experimental $\chi''$($\omega$,$T$) data is analyzed using this scaling form. As seen in the main panel of Fig.~\ref{fig-scal}, a fairly good scaling of the different $\chi''$($\omega$,$T$) curves can be obtained if plotted using the above derived scaling law. The scaling implies $\psi\nu$=0.9 $\pm$ 0.1 and $\tau_0$ = 10$^{-11 \pm 1}$ s, in agreement with the $T_f(f)$ scalings. If we assume that $\nu$ =$-1/\theta$ $\sim$ 3.5 ($\theta$ $\sim$ -0.29\cite{moore}), we obtain $\psi$=0.26, in agreement with theoretical predictions\cite{moore}.

Hence we suggest that as the temperature is lowered below $T_N$, a 2$D$ SG correlation develops in La$_{1.1}$Sr$_{0.9}$MnO$_4$. The 2$D$ character may be related to the $d_{x^2-y^2}$ orbital nature of the electrons injected in the single-layered structure. In this layered case, the interplanar interaction is further weakened by the interplane frustration\cite{onoda,larochelle}, which implies that the AFM state established in LaSrMnO$_4$ and La$_{1.1}$Sr$_{0.9}$MnO$_4$ is quasi 2D. The introduction of additional $e_g$ electrons in the latter introduces the local FM interaction responsible for the 2$D$ SG correlation with Ising anisotropy inherited from the long-ranged AFM state stabilized at higher temperatures. 

To conclude, we have investigated in detail the ac-susceptibility of the electron-doped La$_{1.1}$Sr$_{0.9}$MnO$_4$ single-layered manganite. We have evidenced the appearance of an uncommon two-dimensional spin-glass correlation at low temperatures, as a result of the appearance of ferromagnetic interaction in the quasi  two-dimensional antiferromagnetic state of LaSrMnO$_4$ promoted by the electron doping. Using simple scaling arguments of the droplet model, we have derived a scaling form describing the dynamical scaling of the ac-susceptibility of such a two-dimensional spin-glass state. A good scaling of the experimental data is obtained using this scaling form, confirming the two-dimensional nature of the low-temperature spin-glass state of La$_{1.1}$Sr$_{0.9}$MnO$_4$. Because of great simplifications in two dimension, the scaling form involves only two free parameters, so that the scaling of our experimental data convincingly confirm the validity of the scaling form, and thus the logarithmic growth law of the droplet model.

We thank  Dr. P. E. J\"onsson, and Profs. P. Nordblad and H. Kawamura for valuable discussions.


\begin{thebibliography}{1}
\bibitem{note2}
One such scaling form has been proposed in Ref.~\onlinecite{mydosh}. However, the steps yielding its derivation are unclear, and involve the critical exponent $\gamma$, which implies a large dependence ($\sim$ $T^3$) on the frequency of the magnitude of the maximum of $\chi''$($\omega$,$T$). Such a large dependence is not observed in our system, nor in the other known 2$D$ systems mentioned in the main text. See \onlinecite{mamiya,per-ac}.
\end{thebibliography}
\end{document}